\documentclass[preprint]{ptephy}

\preprintnumber{arXiv:1403.4364} 

\usepackage{bm,latexsym,amsmath,amssymb,amsfonts,mathrsfs}
\DeclareMathOperator\arctanh{arctanh}

\newcommand*{\D}{{\rm d}}

\begin{document}

\title{Exact black hole solutions in shift symmetric scalar-tensor theories}

\author{\name{Tsutomu Kobayashi}{1,\ast}, and \name{Norihiro Tanahashi}{2,3,\ast}}

\address{\affil{1}{Department of Physics, Rikkyo University, Toshima, Tokyo 175-8501, Japan}
\affil{2}{Kavli Institute for the Physics and Mathematics of the Universe, Todai 
Institutes for Advanced Study, University of Tokyo (WPI), 5-1-5 
Kashiwanoha, Kashiwa, Chiba 277-8583, Japan}
\affil{3}{Department of Applied Mathematics and Theoretical Physics,
University of Cambridge, Wilberforce Road, Cambridge CB3 0WA, UK}
\email{tsutomu@rikkyo.ac.jp (TK); norihiro.tanahashi@ipmu.jp (NT)}}

\begin{abstract}%
We derive a variety of
exact black hole solutions in a subclass
of Horndeski's scalar-tensor theory possessing shift symmetry, $\phi\to\phi+c$,
and reflection symmetry, $\phi\to-\phi$.
The theory admits
two arbitrary functions of $X:=-(\partial\phi)^2/2$, and
our solutions are constructed without specifying the concrete form of
the two functions, implying that
black hole solutions in specific scalar-tensor theories
found in the literature can be extended to a more general class of theories
with shift symmetry.
Our solutions include a black hole in the presence of
an effective cosmological constant, the Nariai spacetime, the Lifshitz black hole, and
other nontrivial solutions, all of which exhibit nonconstant scalar-field profile.
\end{abstract}

\subjectindex{E00, E03}

\maketitle

\section{Introduction}

Modifying general relativity has been one of the
most highlighted fields in gravitational physics in recent years.
Long distance modification of gravity
has been studied extensively
so as to explain the current accelerated expansion of the
Universe (see, {\em e.g.}, Ref.~\cite{mog} for a recent review).
More theoretically, it is interesting to ask the simple question
of whether one can consistently modify general relativity
to accommodate, {\em e.g.}, massive gravitons~\cite{Fierz:1939ix,deRham:2010kj,deRham:2014zqa}.
In many cases, modified theories of gravity can be
described, at least effectively, by adding
an extra scalar degree of freedom
that participates in the dynamics of gravity.
The most general Lagrangian composed of
the metric $g_{\mu\nu}$ and a scalar $\phi$
and having second-order field equations
will therefore be a powerful tool to study
various aspects of modified gravity,
and, interestingly, the theory was already
constructed forty years ago by Horndeski~\cite{Horndeski}.
Recently, the Horndeski theory was rediscovered~\cite{Charmousis:2011bf,Deffayet:2011gz}
and shown to be equivalent to
the generalized galileon~\cite{Kobayashi:2011nu}.
While considerable attention has been devoted to
cosmological applications of the Horndeski theory,
black holes in that theory have been less explored so far.

In the context of scalar-tensor modification of gravity,
one of the central questions to address is whether or not
black holes can have scalar hair.
It has been proven by Hawking that a black hole
cannot have scalar hair in the Brans-Dicke theory~\cite{Hawking:1972qk}.
In the traditional scalar-tensor theory where $\phi$
is nonminimally coupled to gravity,
the scalar-no-hair theorem was formulated in~\cite{Bekenstein:1995un}
(under the assumption of spherical symmetry\footnote{See Ref.~\cite{Herdeiro:2014goa}
for a recent attempt to construct a hairy Kerr solution.}), and
a more general proof was provided recently in~\cite{Sotiriou:2011dz},
while the no-hair theorem for a noncanonical scalar field, {\em i.e.},
k-essence, was given in~\cite{Graham:2014mda}.
It is then natural to ask how those results can be extended to
theories whose Lagrangian contains second derivatives of $\phi$.
Such theories are motivated by the galileon~\cite{Nicolis:2008in},
for which the equation of motion still remains of second order.
For the galileon coupled to gravity,
Hui and Nicolis have shown that static and spherically symmetric
black holes cannot be surrounded by any nontrivial profile of
the scalar field~\cite{Hui:2012qt}.


The key of the proof of Ref.~\cite{Hui:2012qt}
is shift symmetry of the scalar field,
{\em i.e.}, symmetry under $\phi\to\phi+c$, where $c$ is a constant,
and the regularity of the square of the Noether current associated
with this symmetry.
Therefore, the same argument seems to hold for more general scalar-tensor theories
with the same symmetry, though there are some loopholes.
One loophole can be opened
by abandoning the static configuration of $\phi$
and/or relaxing some asymptotic conditions on the metric and $\phi$,
and hairy black hole solutions have thus been constructed explicitly
in Ref.~\cite{Babichev:2013cya}.
One can also circumvent the no-hair theorem
by tuning the Lagrangian~\cite{Sotiriou:2013qea}.
In light of the former loophole,
exact black hole solutions with scalar hair have been found
in the theory with nonminimal derivative coupling to the Einstein tensor,
$G^{\mu\nu}\partial_\mu\phi\partial_\nu\phi$~\cite{Babichev:2013cya,Rinaldi:2012vy,Anabalon:2013oea,Minamitsuji:2013ura,Bravo-Gaete:2013dca}.

The term $G^{\mu\nu}\partial_\mu\phi\partial_\nu\phi$
has shift symmetry and reflection symmetry, $\phi\to -\phi$.
The goal of this paper is to extend those previous works
to go beyond this particular example,
giving various exact black hole solutions
with scalar hair in a subclass of the Horndeski theory
possessing shift and reflection symmetries.
The theory contains two {\em arbitrary}
functions of $X:=-(\partial\phi)^2/2$, and we
will provide a variety of solutions without specifying the concrete form of
those functions.

The paper is organized as follows.
In the next section, we present the theory and
the black hole ansatz considered in this paper.
In Secs.~3 and~4, we give various exact solutions
with scalar hair. The regularity of our solutions
is discussed in Sec.~5. Finally, we conclude in Sec.~6.

\section{Shift symmetric scalar-tensor theory and black hole ansatz}
We consider a shift symmetric subclass of the Horndeski theory
whose Lagrangian is given by
\begin{eqnarray}
{\cal L}=G_2(X)
+G_4(X)R+G_{4X}\left[\left(\Box\phi\right)^2
-\left(\nabla_\mu\nabla_\nu\phi\right)^2\right],\label{Lagrangian}
\end{eqnarray}
where $G_2$ and $G_4$ are arbitrary functions of $X$,
and $G_{4X}:=\partial G_4/\partial X$.
The most general shift symmetric scalar-tensor theory
with second-order field equations can accommodate
two more arbitrary functions of $X$, often denoted as $G_3(X)$ and $G_5(X)$
in the literature.
However, we restrict ourselves to the theory
possessing reflection symmetry as well, $\phi\to-\phi$,
which forbids the $G_3$ and $G_5$ terms.
We thus focus on the Lagrangian~(\ref{Lagrangian})
describing the scalar-tensor theory with shift and reflection symmetries.
Since
\begin{eqnarray}
XR+\left(\Box\phi\right)^2
-\left(\nabla_\mu\nabla_\nu\phi\right)^2
=G^{\mu\nu }\partial_\mu\phi\partial_\nu\phi,
\end{eqnarray}
up to a total divergence,
we notice that
the theory considered in
Refs.~\cite{Rinaldi:2012vy,Babichev:2013cya,Anabalon:2013oea,Minamitsuji:2013ura,Bravo-Gaete:2013dca}
corresponds to the specific case with
\begin{eqnarray}
G_2=-2\Lambda_0+2\eta X,
\quad
G_4=\zeta +\beta X,\label{babitheory}
\end{eqnarray}
where $\Lambda_0, \eta, \zeta$, and $\beta$ are constant
parameters. (See Ref.~\cite{Kolyvaris:2011fk,Cisterna:2014nua} for black hole solutions of
the theory~(\ref{babitheory}) in the presence of an electromagnetic field.)
In this paper, we go beyond the specific theory~(\ref{babitheory}),
leaving $G_2(X)$ and $G_4(X)$ arbitrary.
As shown in the following, exact black hole solutions
with a nontrivial configuration of $\phi$
can still be constructed.

Variation with respect to the metric
yields the gravitational field equations,
\begin{eqnarray}
{\cal E}_{\mu\nu}:=
\frac{2}{\sqrt{-g}}\frac{\delta\left(\sqrt{-g}{\cal L}\right)}{\delta g^{\mu\nu}} =0.
\end{eqnarray}
Shift symmetry of the theory allows us to write the
scalar-field equation of motion as a current conservation equation,
\begin{eqnarray}
\nabla_\mu J^\mu=0,\label{seom}
\end{eqnarray}
where
\begin{eqnarray}
J^\mu&:=&-G_{2X}\nabla^\mu\phi +2G_{4X}G^{\mu\nu}\nabla_\nu\phi
-G_{4XX}\left[\left(\Box\phi\right)^2
-\left(\nabla_\mu\nabla_\nu\phi\right)^2\right]\nabla^\mu\phi
\nonumber\\&&
-2G_{4XX}\left(\Box\phi\nabla^\mu X-\nabla^\mu\nabla^\nu\phi\nabla_\nu X\right).
\end{eqnarray}

The metric we are going to study is of the form
\begin{eqnarray}
\D s^2=-h(r)\D t^2+\frac{\D r^2}{f(r)}+r^2\D\Omega^2_K,
\end{eqnarray}
where $\D\Omega^2_K$ is the metric of a unit two-dimensional
sphere, plane, or hyperboloid for $K=+1, 0, -1$, respectively.
We take the following $t$-dependent ansatz
for the scalar field~\cite{Babichev:2013cya},
\begin{eqnarray}
\phi(t, r)=qt+\psi(r),\quad q={\rm const},\label{phi-ansatz}
\end{eqnarray}
for which
\begin{eqnarray}
X=\frac{1}{2}\left[\frac{q^2}{h}-f\left(\psi'\right)^2\right]
\label{X-psi}
\end{eqnarray}
is, however, independent of $t$.
Here and hereafter a prime denotes differentiation with respect to $r$.
Since $G_2$ and $G_4$ are the functions of $X$ only,
it is more convenient to use $X$ rather than $\psi'$
when writing the field equations.
Note in passing that, if $q\neq 0$, $q$ can be chosen arbitrarily
by rescaling the time coordinate:
$q\to q/\alpha$, $t\to\alpha t$, and $h\to h/\alpha^2$ with constant $\alpha$.
This rescaling keeps the line element and $X$ invariant.

\section{$q\neq 0$ solutions}

We will closely follow Ref.~\cite{Babichev:2013cya}
to construct black hole solutions dressed with a time-dependent scalar, $q\neq 0$.
An explicit calculation
using the ansatz~(\ref{phi-ansatz})
shows that
\begin{eqnarray}
J^r&=&\frac{f\psi'}{r^2h}\left\{
-\left(r^2G_{2X}+2KG_{4X}\right)h
+2\left[\left(G_{4X}+2XG_{4XX}\right)\left(rh\right)'-q^2G_{4XX}\right]f\right\},
\,\,\,\,\,\,\\
J^t&=&-\frac{qJ^r}{fh\psi'}-\frac{2q}{r}\left[
G_{4X}\left(\frac{f}{h}\right)'+2G_{4X}'\frac{f}{h}\right],
\label{Jt}
\,\,\,\,\,\,
\end{eqnarray}
and $J^\Omega =0$, where $J^\Omega$
stands for the other two components of the shift current $J^\mu$.
Since the $(t,r)$-component of the gravitational field equations
can be written as
\begin{eqnarray}
0={\cal E}_{tr}=\frac{q}{f}J^r,
\end{eqnarray}
this equation and the scalar-field equation of motion~(\ref{seom})
are satisfied by imposing
\begin{eqnarray}
f=\frac{1}{2}\frac{\left(r^2G_{2X}+2KG_{4X}\right)h}{(G_{4X}+2XG_{4XX})(rh)'-q^2 G_{4XX}}.
\label{solf}
\end{eqnarray}
Now one can see from Eq.~(\ref{Jt}) that $J_\mu J^\mu = -h (J^t)^2$
does not diverge at the horizon as long as $(f/h)'<\infty$.

Using Eq.~(\ref{solf}), we find
\begin{eqnarray}
{\cal E}_{rr}=-\frac{8X\left(G_{4X}^2+G_4G_{4XX}\right)}{r^2\left(r^2G_{2X}+2KG_{4X}\right)h}
\left[\left(K-r^2F\right)(rh)'-\frac{q^2}{2X}\left(K-r^2\Lambda\right)
\right],
\end{eqnarray}
implying that
\begin{eqnarray}
\left[K-r^2F(X)\right](rh)'=\frac{q^2}{2X}\left[K-r^2\Lambda(X)\right].
\label{dh}
\end{eqnarray}
Here we have defined the functions of $X$ as
\begin{eqnarray}
\Lambda(X)&:=&-\frac{1}{2}\frac{G_{2}G_{4XX}+G_{2X}G_{4X}}{G_{4X}^2+G_4G_{4XX}},
\\
F(X)&:=&\frac{G_{2X}G_4-G_2G_{4X}}
{4X\left(G_{4X}^2+G_4G_{4XX}\right)}+\Lambda(X)
\nonumber\\
&=&\frac{\partial_X\left(G_2{\cal G}\right)}{8X\left(G_{4X}^2+G_4G_{4XX}\right)},
\end{eqnarray}
where
\begin{eqnarray}
{\cal G}(X):=2\left(G_4-2XG_{4X}\right).
\end{eqnarray}
Combining Eqs.~(\ref{solf}) and~(\ref{dh}),
one notices the relation
\begin{eqnarray}
f=\frac{2X}{q^2}\left[K-r^2F(X)\right]h.\label{rel-h-f}
\end{eqnarray}
One can exploit the same set of equations, (\ref{solf}) and~(\ref{dh}),
to write the $(t,t)$-component of the gravitational field equations
simply as
\begin{eqnarray}
{\cal E}_{tt}=
-\frac{2h^2}{q^2r}\frac{1}{\cal G}\frac{\D}{\D r}
\left[X{\cal G}^2\left(K-r^2 F\right)\right]=0.\label{ett}
\end{eqnarray}
This equation can be integrated to give
\begin{eqnarray}
X{\cal G}^2(X)\left[K-r^2 F(X)\right]=C,\label{int-ett}
\end{eqnarray}
where $C$ is an integration constant.
Equation~(\ref{int-ett}) determines $X=X(r)$ {\em algebraically}.
Then,
one can integrate Eq.~(\ref{dh}) to determine $h(r)$, and finally
obtain $f(r)$ from Eq.~(\ref{solf}), or, more straightforwardly from Eq.~(\ref{rel-h-f}),
for given $G_2(X)$ and $G_4(X)$.
The angular equations are written as
${\cal E}_{\Omega\Omega} =g_{\Omega\Omega}{\cal E}_{tt}rh'/4h^2=0$
using the equations derived above,
and hence are fulfilled automatically.


\subsection{$F=0$ solutions}
A particularly simple solution of Eq.~(\ref{int-ett})
is obtained if $C$ is chosen in
such a way that
\begin{eqnarray}
C=C_0:= X_F{\cal G}^2(X_F)K,
\end{eqnarray}
where $X_F$ is a {\em constant} satisfying
\begin{eqnarray}
F(X_F)=0.
\end{eqnarray}
In this case, $X=X_F$ trivially solves Eq.~(\ref{int-ett}).
Then, assuming that $K\neq 0$
one can integrate Eq.~(\ref{dh}) immediately to get
\begin{eqnarray}
h=-\frac{\mu}{r}+\frac{q^2}{2X_FK}\left(K-\frac{\Lambda(X_F)}{3}r^2\right),
\end{eqnarray}
where $\mu$ is an integration constant.
(For $K = 0$, Eq.~(\ref{dh}) cannot be satisfied in general.)
Substituting $F(X_F)=0$ to Eq.~(\ref{rel-h-f}), we arrive at
\begin{eqnarray}
f=\frac{2X_FK}{q^2}h,
\end{eqnarray}
and therefore $X_F$ must be such that $2X_FK>0$
for the metric to be Lorentzian.
We can then rescale the time coordinate to set $q^2=2X_FK$.
The final form of the solution is thus
\begin{eqnarray}
f=h=K-\frac{\Lambda(X_F)}{3}r^2-\frac{\mu}{r}.
\end{eqnarray}
This is identical to the black hole metric in the presence
of the (effective) cosmological constant $\Lambda (X_F)$,
though it exhibits a nontrivial profile of $\phi(t,r)$.
Interestingly, $\Lambda(X_F)$ can be nonzero even
in the case where the true cosmological constant
(which could be included in $G_2$) vanishes.
The same metric with an effective cosmological constant
has been constructed
earlier in Ref.~\cite{Babichev:2013cya} for the theory~(\ref{babitheory}).

The radial profile of the scalar field for the above solution
is given by integrating
\begin{eqnarray}
(\psi')^2=\frac{2X_F(1-h)}{fh}.
\end{eqnarray}
In order for $\psi'$ to be real, we must require that
$X_F(1-h)\ge 0$.
This prohibits negative $\Lambda(X_F)$ (with $\mu>0$).
Since $\psi$ diverges as $f, h\to 0$,
one would be concerned about the regularity of the
scalar field at the horizon.
This issue will be discussed in detail in Sec.~\ref{Sec:regularity}.
However, we would emphasize here that
the action depends on the derivatives of the scalar field
rather than $\phi$ itself, and we see that
the coordinate scalar quantities constructed from derivatives, such as
$X$ and $J_\mu J^\mu$, are regular everywhere.

\subsection{Nariai limit of the $F=0$ solutions}
Since the de Sitter-Schwarzschild metric solves the field equations
as shown above,
one can consider its extremal limit, {\em i.e.}, the Nariai spacetime~\cite{Nariai}.
In this Nariai limit, the spacetime between the black-hole and cosmological horizons
can be described by
\begin{eqnarray}
\D s^2=-\left(1-\frac{z^2}{a^2}\right)\D\tilde t^2
+\frac{\D z^2}{\left(1-z^2/a^2\right)}+a^2\D\Omega^2,
\end{eqnarray}
where $a:=1/\sqrt{\Lambda(X_F)}$ and the horizons are located at $z=\pm a$.
The new coordinates $\tilde t$ and $z$ are related to the original ones by
$\tilde t = \epsilon t$ and $z=\epsilon^{-1}(r-a)$, where
the limit $\epsilon\to 0$ is taken keeping
$\tilde t$ and $z$ fixed. 
In these coordinates, the scalar field is written as
\begin{eqnarray}
\phi=
\tilde q
\left[\tilde t\pm a\arctanh(z/a)\right],\label{nariai-scalar}
\end{eqnarray}
where $\tilde q = q/\epsilon$ is kept fixed in the $\epsilon\to 0$ limit.
Even though $\phi$ shows a nontrivial profile and
in particular $|\phi|\to\infty$ as $z\to\pm a$,
we see that $X_F = 0$ and $J_\mu J^\mu=0$ everywhere.
We will discuss
the regularity at the horizons in Sec.~\ref{Sec:regularity}.

\subsection{${\cal G}=0$ solutions}
Another class of $X=$ const solutions
satisfying Eq.~(\ref{ett})
can be obtained
by choosing $X=X_{\cal G}$, where ${\cal G}(X_{\cal G})=0$.
This gives the solution of the form
\begin{eqnarray}
h=\frac{q^2}{2X_{\cal G}}\left[\frac{\Lambda(X_{\cal G})}{F(X_{\cal G})}
+\left(1-\frac{\Lambda(X_{\cal G})}{F(X_{\cal G})}\right)
{\cal T}_\kappa(r)
\right]
-\frac{\mu}{r},\label{newsol}
\end{eqnarray}
where
\begin{eqnarray}
{\cal T}_\kappa(r):=
\begin{cases}
\displaystyle{
\frac{\sqrt{\kappa}}{2r}\ln\left|\frac{r+\sqrt{\kappa}}{r-\sqrt{\kappa}}\right|
} &(\kappa>0)\\
0 &(\kappa=0)\;\;,\\
\displaystyle{
\frac{\arctan\left(r/\sqrt{-\kappa}\right)}{r/\sqrt{-\kappa}}
} &(\kappa<0)
\end{cases}
\end{eqnarray}
with $\kappa:=K/F(X_{\cal G})$.
We find $f=(2X_{\cal G}F(X_{\cal G})/q^2)(\kappa-r^2)h$ by using Eq.~(\ref{rel-h-f}),
and then $\psi$ by integrating Eq.~(\ref{X-psi}).

Let us investigate the solution~(\ref{newsol}) closely
for the three respective cases, $\kappa>0$, $\kappa=0$, and $\kappa<0$.
If $\kappa>0$, $\Lambda/F\neq 1$, and $\mu>0$,
curvature singularities occur at $r=0$ and $r=\sqrt{\kappa}$.
However, those singularities can be hidden behind horizons.\footnote{Note,
however, that the propagation speeds of gravitational waves
and a scalar-field fluctuation can be superluminal in the Horndeski theory.}
This is indeed the case if $h(r)$ has two roots in the interval
$0<r<\sqrt{\kappa}$, 
which is realized for
$X_{\cal G}>0, \quad\Lambda>F>0,\quad K=+1$, 
with not too large $\mu> 0$. 
The geometry between the two zeros
is similar to de Sitter-Schwartzschild.
In the case of $\kappa>0$,
one can also consider the region $r>\sqrt{\kappa}$, with
the singularity at $r=\sqrt{\kappa}$ being hidden inside the horizon.
Such solutions with $h(r)>0$ outside the horizon are obtained
under either of the following conditions:
$X_{\cal G}>0$, $F<\Lambda <0$, $K=-1$ with sufficiently large $\mu$;
$X_{\cal G}>0$, $\Lambda <F<0$, $K=-1$ with arbitrary $\mu$;
$X_{\cal G}<0$, $\Lambda <0<F$, $K=+1$ with arbitrary $\mu$.

For $\kappa=0$ and $\mu>0$,
$f$ and $h$ are positive outside the horizon provided that
$\Lambda<0$ and $X_{\cal G}F<0$.
In this case we are allowed to set $q^2=2X_{\cal G}F/\Lambda$ and
the solution~(\ref{newsol}) reduces simply to
\begin{eqnarray}
f=-\Lambda r^2\left(1-\frac{\mu}{r}\right),
\quad 
h=1-\frac{\mu}{r}.\label{nonafsol}
\end{eqnarray}

If $\kappa<0$ and $\mu\neq 0$, a curvature singularity occurs only at $r=0$.
A solution for which $f, h>0$ everywhere outside the horizon can be obtained
if, for example, $\mu > 0$, $\Lambda<0$, and $X_{\cal G}F<0$.
In the case of $\mu=0$ the solution is regular at the origin.
For example, for $X_{\cal G}>0$, $F<0$, and $\Lambda >0$,
$f$ and $h$ are positive only in a finite region around the origin,
and the geometry is similar to the static region of the de Sitter spacetime.

Finally, 
the special case with $\mu>0$, $X_{\cal G}>0$, $\Lambda(X_{\cal G})=F(X_{\cal G})>0$, $K=+1$
(and hence $\kappa>0$)
reproduces the Schwarzschild black hole in an Einstein static universe and is
discussed in Ref.~\cite{Babichev:2013cya}.


\subsection{Stealth Schwarzschild in the $G_2=0$ theory}
Let us take a look at the theory with $G_2=0$. In this case, we have
$F(X)=\Lambda(X)=0$ for any $G_4$, and
hence Eq.~(\ref{int-ett}) admits {\em only} the $X=$ const solutions.
Taking $q^2=2X$, from Eqs.~(\ref{dh}) and~(\ref{rel-h-f})
we obtain $f=h=1-\mu/r$ for $K=1$.
Therefore, a stealth Schwarzschild black hole solution
can be obtained for more general $G_4$ than taken in~\cite{Babichev:2013cya}
provided that $G_2=0$. For
$K=0$ ($K=-1$),
Eq.~(\ref{solf}) implies that $f=0$ ($f<0$), and therefore
we do not have a sensible solution unless $K=1$.

\section{$q=0$ solutions}

For $q=0$, Eqs.~(\ref{solf}) and~(\ref{dh}) reduce, respectively, to
\begin{eqnarray}
f=\frac{r^2G_{2X}+2KG_{4X}}{G_{4X}+2XG_{4XX}}\frac{h}{2(rh)'},
\label{q0f}
\end{eqnarray}
and
\begin{eqnarray}
K-r^2F(X)=0,\label{Kr2F}
\end{eqnarray}
where $(rh)'\neq 0$ is assumed.
The second equation determines $X=X(r)$ algebraically.
The $(t,t)$-component of the field equations for $q=0$ is given by
\begin{eqnarray}
{\cal E}_{tt}=\frac{h}{r^2}\left[r^2G_2+2KG_4-\frac{1}{{\cal G}}
\frac{\D}{\D r}\left({\cal G}^2 rf\right)\right]=0.
\label{q0Ett}
\end{eqnarray}
We obtain $f(r)$ by integrating Eq.~(\ref{q0Ett}). Then,
Eq.~(\ref{q0f}) is used to fix $h(r)$.

Equation~(\ref{Kr2F}) implies that
$X$ must be dependent on $r$
for $K\neq 0$. In this case, the explicit form of the metric
is dependent on the concrete
form of $G_2$ and $G_4$.
For example, in the theory~(\ref{babitheory}),
Eq.~(\ref{Kr2F}) yields a linear equation in $X$,
which can be solved to give the solution presented in
Refs.~\cite{Rinaldi:2012vy, Babichev:2013cya, Minamitsuji:2013ura}.
For $K=0$, however, Eq.~(\ref{Kr2F}) forces $X$ to be constant, $X=X_F$.
Then, from Eqs.~(\ref{q0f}) and~(\ref{q0Ett}) we obtain
\begin{eqnarray}
f=h=\frac{G_2(X_F)}{3{\cal G}(X_F)}r^2-\frac{\mu}{r},
\end{eqnarray}
where we used
$\partial_X\left(G_2{\cal G}\right)|_{X=X_F}\propto F(X_F)=0$.
This is a planar anti-de Sitter
black hole metric (for $G_2(X_F)/{\cal G}(X_F)>0$).
The profile of the scalar field is given by
$(\psi')^2=-2X_F/f$, implying that $\psi\sim \ln r\to \infty$ as $r\to\infty$,
though its derivative is finite.

Let us finally consider the special class of $X=X_F=$ const
solutions with $q=0$, $K=0$ satisfying
\begin{eqnarray}
G_2(X_F)={\cal G}(X_F)=0.
\end{eqnarray}
(This condition is consistent with $F(X_F)=0$.)
In such theories,
Eq.~(\ref{q0Ett}) is trivially fulfilled.
One can then take {\em any} $f$ and $h$ provided that the two functions satisfy
Eq.~(\ref{q0f}).
For example, Eq.~(\ref{q0f}) admits the following solution:
\begin{eqnarray}
f=r^2\left(1-\frac{\mu}{r^{n }}\right),
\quad
h=r^{b-1}\left(1-\frac{\mu}{r^n}\right)^{b/n},
\end{eqnarray}
where $b:=-G_{2X}/{\cal G}_{X}|_{X=X_F}$ and $n$ is arbitrary.
A particular case $b=n \,(=2z+1)$
corresponds to the Lifshitz black hole solution
with the dynamical exponent $z$~\cite{Bravo-Gaete:2013dca}, though it was originally
constructed in the specific theory~(\ref{babitheory}).
As another example, it is also easy to check that
\begin{eqnarray}
f=b r^2\left(1-\frac{\mu}{r}\right),
\quad
h=1-\frac{\mu}{r},
\end{eqnarray}
fulfill Eq.~(\ref{q0f}).
This solution gives the same geometry as obtained from Eq.~(\ref{nonafsol}).

\section{Regularity at the horizon?}
\label{Sec:regularity}

We are now in position to check the regularity of
the scalar field at the horizon, $r=r_h$.
Suppose that $f$ and $h$ are expanded near the horizon as
\begin{eqnarray}
h=h_1(r-r_h)+\cdots,\quad f=f_1(r-r_h)+\cdots,
\label{expandfh}
\end{eqnarray}
where $f_1$ and $h_1$ are constants.
So far we have concentrated mainly on $X=$ const solutions.
It turns out, however, that
the regularity at the horizon does not depend on this property.
One can determine $X$ from Eq.~(\ref{int-ett}) ($q\neq 0$)
or Eq.~(\ref{Kr2F}) ($q=0$), and
$X$ thus obtained may be constant or may be $r$-dependent.
In any case, since those algebraic equations do
not depend on the metric functions explicitly,
nothing special happens to $X$ at the horizon.
This fact allows us to write
\begin{eqnarray}
X=X_h+{\cal O}(r-r_h),
\label{expandX}
\end{eqnarray}
where the constant $X_h$ is fixed by
$X_h{\cal G}^2(X_h)[1-r_h^2F(X_h)]=C$ for $q\neq 0$
and $K-r_h^2F(X_h)=0$ for $q=0$.
Substituting Eqs.~(\ref{expandfh}) and (\ref{expandX}) to Eq.~(\ref{Jt}),
one can see that $J_\mu J^\mu$ is also finite at the horizon.
Thus,
all the solutions we have found in this paper are regular in the sense that 
not only the spacetime but also the coordinate invariants
constructed from the derivatives of
scalar field, $X$ and $J_\mu J^\mu$, are regular at the horizon.

Let us then examine the behavior of $\phi$ itself near the horizon,
though the action depends on the scalar through $\partial_\mu\phi$
due to shift symmetry.
From Eq.~(\ref{X-psi}), it is found in the $q\neq 0$ case that
\begin{eqnarray}
&&\psi'=\pm\frac{q}{\sqrt{f_1h_1}}\frac{1}{r-r_h}
+{\cal O}\bigl((r-r_h)^0\bigr)
\nonumber\\
\Rightarrow&&
\!
\psi=\pm\frac{q}{\sqrt{f_1h_1}}\ln\left|\frac{r-r_h}{r_h}\right|
+{\cal O}\bigl((r-r_h)^0\bigr).
\label{psi-sol}
\end{eqnarray}
Apparently, $\phi$ diverges logarithmically as $r\to r_h$. However,
as in~\cite{Babichev:2013cya}, one may
introduce the ingoing Eddington--Finkelstein coordinates $(v, r)$ defined by
$\D v=\D t + \D r/\sqrt{fh}$
to write
$\phi\simeq qv$ near the horizon,
where the plus sign was chosen in Eq.~(\ref{psi-sol}).
Thus, we see that the scalar field
is in fact regular at $r=r_h$.
Note that if the solution exhibits a cosmological horizon
in addition to the black hole horizon at $r=r_h$ then it will correspond to $v=\infty$
and hence $\phi$ is no longer regular there.
In any case, it should be emphasized again that
the coordinate invariants constructed from $\phi$'s derivatives are regular,
as we noted above.
The argument here also applies to the $F=0$ solutions in the Nariai limit
by replacing $r$ with $z$.


In the $q=0$ case, it is easy to see $\phi \simeq 2[(-2X_h/f_1)(r-r_h)]^{1/2}$
near the horizon, and hence $\phi$ is regular at the horizon.
As seen in the previous section, $\phi\to\ln r$ as $r\to\infty$
for the solutions in this class.

\section{Conclusions}
In this paper, we have derived a variety of exact black hole solutions
in a subclass of Horndeski's scalar-tensor
theory having shift and reflection symmetries.
Assuming the time-dependent ansatz for the scalar field~\cite{Babichev:2013cya},
$\phi(t,r)=qt+\psi(r)$, and following the method of constructing solutions
of Ref.~\cite{Babichev:2013cya},
we have obtained a solution describing a black hole in
the presence of a cosmological constant and other nontrivial
solutions without fixing the concrete form of
the two arbitrary functions in the theory.
This was made possible because the solutions we have explored
have the key property: $X=$ const.
Our solutions circumvent
the no-hair theorem because
the scalar field itself is not static
or it diverges at infinity.
Note that the action depends only on the derivatives of $\phi$ due
to shift symmetry, and as a consequence
the spacetime is static
and the coordinate invariants constructed from the
derivatives of the scalar field are regular.

There are a lot of remaining issues to be addressed.
One of the basic questions regarding a black hole solution is
whether or not it is stable.
It would therefore be interesting to study the stability issue
by extending
the black hole perturbation theory in Horndeski's scalar-tensor gravity
formulated in the static background~\cite{Kobayashi:2012kh}.
It would also be worth trying to generalize our solutions
to rotating ones as has been done for 
a conformal scalar field~\cite{Bhattacharya:2013hvm}, or
to black holes with a realistic matter distribution, as
in Ref.~\cite{Davis:2014tea}
for typical modified gravity theories with a screening mechanism.
We primarily focused on $X=$ const solutions in this paper,
even though there could be many interesting solutions with
$r$-dependent $X$. 
It would be useful to clarify the phase space of possible 
solutions and to discuss their phenomenological properties 
as in Refs.~\cite{Kaloper:2011qc,Sbisa:2012zk,Tasinato:2013rza,Kaloper:2013vta}.
The theory we have studied admits 
superluminal propagations. It is crucial to study whether such superluminal propagations
appear around the present black hole background in order to
understand their causal structures
and the influence of singularities on the region exterior to the horizon.

Finally, we would like to remark that
the Euclidean version of the theory~(\ref{Lagrangian})
has been used in constructing a
mechanism of emergent Lorentz signature~\cite{Mukohyama:2013ew}.
Our black hole solutions may help to give an insight about
such a novel scenario.

\ack
We would like to thank Christos~Charmousis and Eugeny~Babichev
for pointing out an erroneous statement in the first version of the paper.
We also thank Hayato~Motohashi for a comment
on the earlier draft.
This work was supported in part by JSPS Grant-in-Aid for Young
Scientists (B) No.~24740161 (T.K.).  
The work of N.T. is supported in part by World Premier International 
Research Center Initiative (WPI Initiative), MEXT, Japan, 
and JSPS Grant-in-Aid for Scientific Research 25$\cdot$755. 




\end{document}